\documentclass[letterpaper, 10 pt, conference]{ieeeconf}  

\IEEEoverridecommandlockouts                              

\usepackage{amsfonts}
\usepackage{mathrsfs}
\usepackage{amssymb}
\usepackage{extarrows}
\usepackage{color}
\usepackage{graphicx}
\usepackage{cite}

\newtheorem{theorem}{Theorem}
\newtheorem{lemma}{Lemma}

\newtheorem{proposition}{Proposition}

\def\qed{ \rule{.08in}{.08in}}

\title{A Generalized Discrete-Time Altafini Model}

\author{L. Wang$^{1}$,   J. Liu$^{2}$, A. S. Morse$^{1}$, B. D. O. Anderson$^3$, and  D. Fullmer$^{1}$
\thanks{This work was  supported by  NSF grant  1607101.00,
  AFOSR grant  FA9550-16-1-0290, and ARO grant  W911NF-17-1-0499}
\thanks{$^{1}$ L. Wang, A. S. Morse and D. Fullmer are with the Department
of Electrical Engineering, Yale University, New Haven, CT, USA.
        {\tt\small \{lili.wang,  as.morse, daniel.fullmer\}@yale.edu }}%
\thanks{$^{2}$ J. Liu  is with Stony Brook University, Stony Brook, New York
        {\tt\small ji.liu@stonybrook.edu }}%
\thanks{$^{3}$ B. D. O. Anderson is with  Research School of Engineering, Australian National University, Action ACT, Australia
        {\tt\small Brian.Anderson@anu.edu.au }}%
        }
\begin{document}

 \maketitle
 \thispagestyle{empty}

\begin{abstract}
A discrete-time modulus consensus model is considered in which the
interaction among a family of networked agents is described by a
time-dependent gain graph whose vertices correspond to agents and
whose arcs are assigned complex numbers from a cyclic group.
Limiting behavior of the model is studied using a graphical
approach. It is shown that, under appropriate connectedness, a
certain type of clustering will be reached exponentially fast for
almost all initial conditions if and only if the sequence of gain
graphs is ``repeatedly jointly structurally balanced'' corresponding
to that type of clustering, where the number of clusters is at most
the order of a cyclic group. It is also shown that the model will
 reach a consensus asymptotically at zero if the sequence of gain
graphs is repeatedly jointly strongly connected and structurally
unbalanced. In the special case when the cyclic group is of order
two, the model simplifies to the so-called Altafini model whose gain
graph is simply a signed graph.
\end{abstract}



\section{Introduction}

With the rapid expansion of online social services, there has been
an increasing interest in understanding how individuals' opinions
and behaviors evolve over time in a social network \cite{centola}.
Opinion dynamics has a long history in the social sciences
\cite{french}. Probably the simplest and most well-known model of
opinion dynamics is the classical DeGroot model originated in
statistics \cite{degroot}. The DeGroot model deals with a
time-invariant connected network of individuals, each of which
updates his/her opinion by taking a convex combination of the
opinions of his/her neighbors at each discrete time step. The model
is also called a consensus model and has attracted considerable
attention in the systems and control community 
\cite{Ts3,vicsekmodel,luc,reza1,ReBe05,survey,reachingp1,cut,cdc14,rate},
with a focus on time-varying networks. It is well known that under
appropriate joint connectivity assumptions, the DeGroot model with
time-varying neighbor relationships causes all individuals' opinions
to reach a consensus.

 Although consensus is an important collective phenomenon,
 splits of opinions on issues are often observed in social networks, such as political polarization \cite{delia} and cohesive subgroups \cite{coh}.
 Various models have been proposed for opinion dynamics to understand and explain the formation of polarization, fragmentation, and clustering of opinions in a social network. Notable examples include the Friedkin-Johnsen model \cite{johnsen,magazine}, the Hegselmann-Krause model
 \cite{krauseTAC,etesami}, which explore the effects of individuals' stubbornness and homophily, respectively.
 Specifically, the Friedkin-Johnsen model may lead to fragmentation of opinions, and the Hegselmann-Krause model can cause clustering among the individuals, while the number of clusters is unpredictable.

In recent results, the so-called Altafini model \cite{altafini},
incorporates in the DeGroot model a binary social relationship among
individuals. Specifically, the Altafini model uses a signed,
directed graph to depict the neighbor relationships among the
individuals, in which vertices correspond to individuals, directions
of arcs indicate directions of information flow, and each (directed)
arc is associated with a positive or negative sign in that positive
signs represent friendly or cooperative relationships and negative
signs represent antagonistic or competitive relationships. The
continuous-time Altafini model has been considered in
\cite{altafini,lift,mingtac,cdc15tac},
and the discrete-time counterpart has been studied in
\cite{modulus,lift,weiguo,cdc15tac}. For the discrete-time Altafini
model over time-varying signed directed graphs, it was shown in
\cite{modulus} that for any ``repeated jointly strongly connected''
sequence of graphs, the absolute values of all individuals' opinions
will asymptotically reach a consensus, which has consensus and
two-clustering as special cases.
Necessary and sufficient conditions for exponential convergence with
respect to each possible type of limit state were established in
\cite{cdc15tac} in terms of structural balance/unbalance, a concept
from social sciences \cite{harary}.

The Altafini model is restricted to two clusters. In a realistic
social network, multiple clusters of opinions occur from time to
time. Thus, there is ample motivation to generalize or modify the
Altafini model, which yields the possibility of multiple clusters.
In \cite{group}, a generalization of the continuous-time Altafini
model was proposed by allowing the gains of the neighbor graph to be
the elements of a finite group, with the order of the group
determining the largest possible number of clusters; the paper
considers fixed neighbor graphs. Another generalization was
introduced in \cite{jiugang} which allows the weights to be any
complex numbers. It was shown in \cite{jiugang} that when the
complex-weighted neighbor graph is fixed and strongly connected,
either all individuals' opinions converge to zero, or their
magnitudes reach a consensus, which is called a modulus consensus;
the paper also considers a special discrete-time model. A
discrete-time counterpart of the model in \cite{jiugang} was studied
in \cite{acc17} which studies time-varying graphs and establishes
sufficient conditions for exponential convergence. It is worth
emphasizing that both the models in \cite{jiugang} and \cite{acc17}
require nontrivial matrix analysis to determine the maximum possible
number of clusters.

In this paper, we consider a generalized discrete-time Altafini
model over time-varying directed graphs, in which the gains are
complex numbers from a cyclic group whose order determines the
maximum possible number of clusters. Although the model is a special
case of the model in \cite{acc17}, such a setting allows us to
analyze the model using a graphical approach and establish a
necessary and sufficient condition for exponentially fast nonzero
modulus consensus, whereas only a sufficient condition was provided
in \cite{acc17}. A sufficient condition for asymptotic consensus at
zero is also provided. It turns out that the cyclic group composed
of complex numbers is a special case of the group considered in
\cite{group}. We focus on the cyclic group for simplicity. It
appears likely that the results derived in this paper can be
generalized to any point group, which will be addressed in future
work.

Complex-weighted graphs and the associated complex-valued adjacency
matrices find applications in
formation control \cite{zhiyun,circular} and localization problems
\cite{zhiyun1}.
The work in this paper is also related to ``group consensus''
\cite{wanggroup} and ``cluster synchronization''
\cite{chen09,caogroup,allgower}.



 \section{Problem Formulation}

We are interested in a network of $n$ agents labeled $1,2,\ldots, n$
which are able to receive information from their neighbors where by
the neighbor of agent $i$ is meant any other agent in agent $i$'s
reception range. We write $\mathcal{N}_i(t)$ for the set of labels
of agent $i$'s neighbors at discrete time $t=0,1,2,\ldots$ and we
take agent $i$ to be a neighbor of itself. A directed graph
$\mathbb{G}$ with $n$ vertices labeled $1,2, \ldots, n$ is a
\textit{gain graph} if each arc $(j,i)$ is assigned a \textit{gain}
$g_{ij}$ where $g_{ij}$ is a complex number from the cyclic group $
\mathcal{G}= \{e^{\frac{2\pi(k-1)}{m}j}:k\in \mathbf{m}\}$;  here
$m$ is a positive integer greater than $1$ and
$\mathbf{m}=\{1,2,\cdots, m\}$. We say that $\mathbb{G}$ is a gain
graph associated with the gain set $\mathcal{G}$. The simplest case
of a gain graph is when $m=2$ in which case the set of possible
gains is $\{1,-1\}$ and $\mathbb{G}$ is typically called a
\textit{signed graph} \cite{altafini}. One interpretation for a
signed graph is that agent $i$ is a friend of agent $j$ if arc
$(j,i)$ is assigned with $1$, or a foe of agent $j$ if arc $(j,i)$
is assigned with $-1$. It is more difficult to assign meaning to a
gain graph if $m>2$. Nonetheless such graphs have found applications
in network flow theory,  geometry, and  physics\cite{Zas}.
Neighbor relations at time $t$ are characterized by a gain graph
$\mathbb{N}(t)$ associated with the gain set $\mathcal{G}$ with $n$
vertices, and a set of arcs defined so that there is an arc from
vertex $j$ to vertex $i$ whenever agent $j$ is a neighbor of agent
$i$. It is natural to assume that each self-arc in $\mathbb{N}(t)$
is assigned with a gain ``$1$''.

Each agent $i$ in the network has a complex-valued state $x_i(t)$
and updates its state using a discrete-time iterative rule given by
\begin{equation}\label{eq:update_i}
x_i(t + 1) =\frac{1}{m_i(t)}\sum_{j\in\mathcal {N}_i(t)}
g_{ij}(t)x_j(t)\;\;\;\; t\geq 0
\end{equation}
where $m_i(t)$ is the number of neighbors of agent $i$ at time $t$,
and  $g_{ij}(t)\in \mathcal{G}$ is the gain assigned to the arc
$(j,i)$.

The $n$ update equations in~\eqref{eq:update_i} can be written as
one linear recursion equation
\begin{equation}\label{eq:update_sys}
x(t+1)=G(t)x(t),\;\;\; t\geq 0
\end{equation}
where each $x(t)$ for $t\geq 0$ is a vector in $\mathbb{C}^n$ whose
$i$th entry is $x_i(t)$, and $G(t)$ is an $n\times n$ matrix whose
$ij$th entry is $\frac{1}{m_i(t)}g_{ij}(t)$ if $j\in
\mathcal{N}_i(t)$, or $0$ if $j\notin \mathcal{N}_i(t)$. We call
matrix $G(t)$  a \textit{gain matrix} of the gain graph
$\mathbb{N}(t)$. Let $F(t)$ be the \textit{flocking matrix} of
$\mathbb{N}(t)$ whose $ij$-th entry $f_{ij}(t)$ is
$\frac{1}{m_i(t)}$ if $j\in \mathcal{N}_i(t)$, or $0$ if $j\notin
\mathcal{N}_i(t)$. It is easy to see that $|G(t)| = F(t)$ where
$|G(t)|$ is the $n\times n$ matrix which results when each entry of
$G(t)$ is replaced by its modulus.

We are interested in the convergence of the state in
system~\eqref{eq:update_sys}. System~\eqref{eq:update_i} or
\eqref{eq:update_sys} achieves \textit{modulus consensus} if
\[\lim_{t\rightarrow \infty} |x_i(t)|=
\lim_{t\rightarrow \infty} |x_j(t)|,\;\; \; \forall  i,j \in
\mathbf{n}  \] where $\mathbf{n}= \{1,2,\ldots,n\}$. Moreover,
system~\eqref{eq:update_i} or \eqref{eq:update_sys} achieves
\textit{$m$-modulus consensus} if it achieves modulus consensus and
$\mathbf{n}$ can be partitioned into $m$ subsets
such that for any $i$, $j\in \mathbf{n}$, $\lim_{t\rightarrow
\infty} x_i(t)= \lim_{t\rightarrow \infty} x_j(t)$ if $i$ and $j$
are in the same subset, and $\lim_{t\rightarrow \infty} x_i(t) \neq
\lim_{t\rightarrow \infty} x_j(t)$ if $i$ and $j$ are in different
subsets.

The problem of interest is to derive  necessary and sufficient
graphical conditions on a sequence of $\mathbb{N}(t)$ under which
system \eqref{eq:update_sys} will achieve $m$-modulus consensus
exponentially fast.

\section{Main Results}

In this section, we first introduce some definitions and notation.
Then main results of the paper are given.

Given a gain graph $\mathbb{G}$ associated with gain set
$\mathcal{G}$,  a \textit{walk} in $\mathbb{G}$ is a sequence of
vertices  connected by arcs corresponding to the order of the
vertices in the sequence. We define the\textit{ gain along a walk}
of a gain graph to be the product of gains assigned to arcs in the
walk. A \textit{semi-walk} in $\mathbb{G}$ is a sequence of vertices
connected by arcs in which the arc directions are ignored. We define
the \textit{gain along a semi-walk} of a gain graph to be the value
of  the product of gains assigned to arcs whose directions are
consistent with the order of the vertices in the semi-walk
multiplying the product of the inverse of gains assigned to  arcs
whose directions are consistent with the reverse order of the
vertices in the semi-walk.  A walk is called a \emph{path} if there
is no repetition of vertices in the walk. A walk is \emph{closed} if
it has the same starting vertex and ending vertex. A walk is a
\emph{cycle} if  it is closed and there is no repetition of vertices
in the walk except the starting and ending vertices.
  A semi-walk is called a \emph{semi-path}
if there is no repetition of vertices in the semi-walk. A semi-walk
is \emph{closed} if it has the same starting vertex and ending
vertex. A semi-walk is a \emph{semi-cycle} if  it is closed and
there is no repetition of vertices in the semi-walk except the
starting and ending vertices.
%
%
For the gain graph $\mathbb{G}$ associated with the gain set
$\mathcal{G}$, it is said to be \textit{structurally $m$-balanced}
if all semi-walks joining the same ordered pair of vertices in
$\mathbb{G}$ have the same gain. Otherwise, the gain graph
$\mathbb{G}$ is \textit{structurally unbalanced}.

According to \cite{group},  the following lemma  can be used to
check whether a gain graph associated with the gain set
$\mathcal{G}$ is structurally $m$-balanced or unbalanced.

\begin{lemma} \label{lemma:semi-cycle1}Let $\mathbb{G}$ be a gain graph associated with the gain set $\mathcal{G}$, $\mathbb{G}$ is \textit{structurally $m$-balanced} if and only if all the semi-cycles of $\mathbb{G}$ have gain 1.
If, in addition, $\mathbb{G}$ is strongly connected,  $\mathbb{G}$
is \textit{structurally $m$-balanced} if and only if all the cycles
of $\mathbb{G}$ have gain 1.
\end{lemma}

More can be said.
 Let $\mathcal{I}$ be a set of vectors in $\mathbb{C}^n$ such that
for each $b\in \mathcal{I}$,  $b(1)=1$ and $b(i)$ is an element in
$\mathcal{G}$. Here $b(i)$ is the $i$-th entry of vector $b$ and
$i\in \mathbf{n}$. We call an element $b$ a \textit{clustering
vector}. Consider a gain graph $\mathbb{G}$ with $n$ vertices
labeled $1,2,\ldots,n$ and associated with the gain set
$\mathcal{G}$. The vertex set $\mathbf{n}$ of $\mathbb{G}$ will be
separated into $m$ disjoint sets $V_1,V_2,\ldots, V_m$ such that
$\cup_{i=1}^mV_i=\mathbf{n}$ according to a given clustering vector
$b$. For any vertex $i\in \mathbf{n}$, $i\in V_p$ if
$b(i)=e^{\frac{2\pi (p-1)}{m}j}$ for any $p\in \mathbf{m}$. Set
$1\in V_1$. If  all semi-walks from vertex $i\in V_p$ to vertex
$j\in V_q$ have the same gain $e^{\frac{2\pi (p-q)}{m}j}$ for $p\geq
q\in \mathbf{m}$, the gain graph $\mathbb G$ associated with
$\mathcal{G}$ is said to be \textit{structurally $m$-balanced with
respect to the clustering vector $b$}. Otherwise we say $\mathbb{G}$
is \emph{not structurally $m$-balanced with respect to the
clustering vector $b$}. Later in Section~\ref{sec:analysis} ,we will
show that  any gain graph $\mathbb{G}$ is structurally $m$-balanced
if and only if there exist a clustering vector $b$ such that
$\mathbb{G}$ is structurally $m$-balanced with respect to the vector
$b$.

A finite sequence of directed graphs $\mathbb{G}(1),\mathbb{G}(2),
\ldots, \mathbb{G}(p)$ with the same vertex set is \emph{jointly
strongly connected} if the union\footnote{The union of a finite
sequence of directed graphs with the same vertex set is a directed
graph with the same vertex set and the arc set which is the union of
the arc sets of all directed graphs in the sequence.} of the
directed graphs in this sequence is strongly connected. Meanwhile,
an infinite sequence of directed graphs
$\mathbb{G}(1),\mathbb{G}(2),\ldots$ with the same vertex set is
\emph{repeatedly jointly strongly connected} if there exist positive
integers $p$ and $q$ such that each the finite sequence
$\mathbb{G}(q+kp),\mathbb{G}(q+kp+1),\ldots, \mathbb{G}(q+kp+p-1)$
for all $k\geq 0$ is jointly strongly connected. Based on the above
definitions, we give the definitions of jointly balancedness  for
gain graphs. A finite sequence of directed gain graphs
$\mathbb{G}(1),\mathbb{G}(2),\ldots, \mathbb{G}(p)$ with the same
vertex set and same gain set $\mathcal{G}$ is \emph{jointly
structurally $m$-balanced} with respect to a clustering vector $b\in
\mathcal{I}$ if the union\footnote{The union of a finite sequence of
directed gain graphs with the same vertex set is a "multi-directed
gain graph? which can have multiple (can be more than two) directed
arcs from a vertex i to another vertex j with different gains.} of
the gain graphs in this sequence is structurally $m$-balanced with
respect to the vector $b$. If there is no $b\in \mathcal I$ such
that the union of the digraphs in this sequence is structurally
$m$-balanced with respect to, this sequence of graphs is
\emph{jointly structurally unbalanced}. Meanwhile, an infinite
sequence of  directed gain graphs
$\mathbb{G}(1),\mathbb{G}(2),\ldots$ with the same vertex set and
same gain set $\mathcal{G}$ is \emph{repeatedly jointly structurally
$m$-balanced} with respect to a clustering vector $b\in \mathcal{I}$
(or \emph{repeatedly jointly structurally unbalanced}) if there
exist positive integers $p$ and $q$ such that each finite sequence
$\mathbb{G}(q+kp),\mathbb{G}(q+kp+1),\ldots, \mathbb{G}(q+kp+p-1)$
is structurally $m$-balanced with respect to $b$ (or jointly
structurally unbalanced) for all $k\geq 0$. It is worth emphasizing
that the converse of repeatedly jointly structurally $m$-balanced is
not repeatedly jointly structurally unbalanced.

The main results of this paper are as follows.

\begin{theorem}\label{thm:1}
 Suppose that the sequence
of neighbor graphs
$\mathbb{N}(0),\mathbb{N}(1),\mathbb{N}(2),\ldots$ with the same
gain set $\mathcal{G}$ is repeatedly jointly strongly connected.
System~\eqref{eq:update_sys} reaches an $m$-modulus consensus
corresponding to $b \in \mathcal I$ exponentially fast for almost
all initial conditions if and only if the graph sequence
$\mathbb{N}(0),\mathbb{N}(1),\mathbb{N}(2),\ldots$ is repeatedly
jointly structurally $m$-balanced with respect to the clustering
vector $b$.
\end{theorem}

\begin{theorem}\label{thm:2}
Suppose that the sequence of neighbor graphs
$\mathbb{N}(0),\mathbb{N}(1),\mathbb{N}(2),\ldots$ with the same
gain set $\mathcal{G}$  is repeatedly jointly strongly connected.
System~\eqref{eq:update_sys} asymptotically converges to zero for
all initial conditions if the graph sequence
$\mathbb{N}(0),\mathbb{N}(1),\mathbb{N}(2),\ldots$ is repeatedly
jointly structurally unbalanced.
\end{theorem}

Both theorems are proved in the next section.

\section{Analysis}\label{sec:analysis}

In this section, we first give a result on the graph structurally
$m$-balanceness. Analysis on Theorem~\ref{thm:1} and
Theorem~\ref{thm:2} will be provided as well.

\begin{proposition}\label{prop:set}
Let  $\mathbb{G}$ be a gain graph  with $n$ vertices labeled $1,2,
\ldots, n$ and associated with  gain set $\mathcal{G}$. $\mathbb{G}$
is structurally $m$-balanced if and only if there exist a clustering
vector $b$ such that $\mathbb{G}$ is structurally $m$-balanced with
respect to the vector $b$.
\end{proposition}


 \vspace{0.1in} \noindent {\bf Proof:}
   (Sufficiency) Since there exist a clustering
 vector $b$ such that $\mathbb{G}$ is structurally $m$-balanced with
 respect to the vector $b$, we can get  $m$ disjoint sets
 $V_1,V_2,\ldots, V_m$ such that $\cup_{i=1}^mV_i=\mathbf{n}$,
  $1\in V_1$, and all semi-walks from
 vertex $i\in V_p$ to vertex $j\in V_q$ have the same gain
 $e^{\frac{2\pi (p-q)}{m}j}$ for $p\geq  q\in \mathbf{m}$.
  Since  all semi-walks from vertex $i\in
 V_q$ to $j \in V_p$  have the same gain,  all semi-walks joining the
 same pair of vertices have the same gain. According to the
 definition of structurally $m$-balanced graph, the gain graph
 $\mathbb{G}$ is structurally $m$-balanced.

 (Necessity)
  Start with $m$ empty vertex sets
 $V_1, \ldots, V_m$. First, choose  vertex $1$ to be in set $V_1$. If
 a vertex $i \in  \mathbf{n}$ is disconnected to vertex $1$,
 $i\in V_1$. If a vertex $i \in \mathbf{n}$ is weakly connected
 to  vertex $1$, all the semi-walks between $i$  and vertex $1$
 have gain $1$, $i \in V_1$. Repeat this procedure until there is no
 such vertex which can be found in $ \mathbf{n}$. Next, for any
 vertex $i \in \mathbf{n}/V_1$, if there exists a vertex  in
 $V_1$ such that all the semi-walks from the vertex in $V_1$ to vertex $i$
 have gain $e^{\frac{2\pi (p-1)}{m}j}$, let the vertex $i \in V_{p}$
 where $p\in {\mathbf{m}}/\{1\}$. The $m$ disjoint sets $V_1,\ldots,
 V_m$ have been obtained. Next, we are going to show
 $\cup_{i=1}^mV_i= \mathbf{n}$. It is obvious that $\cup_{i=1}^m
 V_i\subset \mathbf{n}$. On the other hand,   any vertex $v\in
 \mathbf{n}/ V_1$ must be weakly  connected to vertex $1 \in V_1$
 with a gain $e^{\frac{2\pi( q-1)}{m}j}$ where $q\in
 \mathbf{m}/\{1\}$. Since graph $\mathbb{G}$ is structurally
 $m$-balanced, all semi-walks with the same starting and ending
 vertices have the same gain. Then $v\in V_{q}$. Thus, $
 \mathbf{n}\subset \cup_{i=1}^mV_i$. In all, $ \cup_{i=1}^mV_i=
 \mathbf{n}$.

    Suppose there is a semi-walk
 from vertex $i\in V_q$ to $j \in V_p$ where $p\geq  q $ whose gain
 is  $g^*$. The gain of a semi-walk from a vertex $m_1\in V_1$ to $i$
 is $e^{\frac{2\pi (q-1)}{m}j}$ while the gain of a semi-walk from
 vertex $m_2\in V_1$ to $j$ is $e^{\frac{2\pi (p-1)}{m}j}$. Since
 there is a semi-walk from vertex $i$ to $j$ with a gain $\alpha^*$,
 there is a semi-walk from vertex $m_1$ to $m_2$ with gain
 $e^{\frac{2\pi (q-1)}{m}j}e^{-\frac{2\pi (p-1)}{m}j}g^*$  which is
 $e^{\frac{2\pi (q-p)}{m}j}g^*$. As defined earlier in the proof
 $e^{\frac{2\pi (q-p)}{m}j}g^*=1$. Thus  $g^*= e^{\frac{2\pi
 (p-q)}{m}j}$.  In all, all semi-walks from vertex $i\in V_q$ to $j
 \in V_p$ where $p\neq q $ have the same gain which is  equal to
 $e^{\frac{2\pi (p-q)}{m}j}$ for $p\geq q$. For $i\in \mathbf{n}$,
 let $b(i)=e^{\frac{2\pi (p-1)}{m}j}$ if  $i\in V_p$ for $p\in
 \mathbf{m}$.  This clustering vector $b$ is the vector such that
 $\mathbb{G}$ is structurally $m$-balanced with respect to. \hfill
 $\qed$ \vspace{0.1in}

When $m=2$, the model becomes the Altafini model which has been well
studied in \cite{altafini,modulus,weiguo,cdc15tac}. As defined in
\cite{altafini}, a graph $\mathbb{G}$ with the gain set $\{1, -1\}$
is structurally $2$-balanced if the vertices of $\mathbb{G}$ can be
partitioned into two sets such that each arc connecting two agents
in the same set has a positive gain and each arc connecting two
agents in different sets has a negative gain. This definition also
satisfies Proposition~\ref{prop:set} for which we can say the graph
is structurally $2$-balanced with respect to a clustering vector
composed of $1$ and $-1$.

In the following, we are going to show each element $b$ in
$\mathcal{I}$ uniquely defines a clustering pattern of all the
agents in the connected network by the gains of the entries of $b$.
That is if two entries say $b(i)$ and $b(j)$ of vector $b$ have the
same gain, agents $i$ and $j$ are in the same clustering. For a
structurally $m$-balanced graph $\mathbb G$  with gain set
$\mathcal{G}$, the agents in the same vertex set $V_i$  for $i\in
\mathbf{m}$ will converge to the same value.

Define a time-dependent $mn$-dimensional vector $z(t)$ such that for
each time $t$,
\[z(t)=
\begin{bmatrix}
\alpha_0 x(t)\\ \alpha_1 x(t)\\ \vdots\\ \alpha_{m-1} x(t)
\end{bmatrix}
\]
where $\alpha_i=e^{\frac{2\pi i}{m}\iota}$  for $i\in
\underline{\mathbf {m} }=\{0,1,2,\ldots,m-1\}$.

Then for all $i\in \{1,2\ldots, mn\}$,
\[
z_i(t+1)=\sum_{j=1}^{mn}\bar{a}_{ij}(t)z_j(t)
\]
in which    if $g_{ij}(t)=1$  for $ p\in \underline{\mathbf{m}}$ and
$i,j \in \mathbf{n}$
\[
\bar{a}_{i+pn,j+pn}=\max \{f_{ij}(t),0 \} ,\]if $g_{ij}(t)=\alpha_q
$ for a fixed $ q\in \underline{\mathbf{m}}$, for each $\ p\in
\underline{\mathbf{m}}$, and $i,j \in\mathbf{n}$,
\[\bar{a}_{i+pn, j+((p+q)\text{ mod }m)n} =\max
\{f_{ij}(t),0 \}\]where $(p+q)\text{ mod }m$ is the remainder of
$p+q$ divided by $m$. It is obvious that the expanded system is
equivalent to system~\eqref{eq:update_sys}. The system can be
written in the form of a state equation
\begin{equation}\label{eq:expand_state}
z(t+1)=\bar{G}(t)z(t)
\end{equation}
where $\bar{G}(t)=[\bar{a}_{ij}(t)]$ is an $mn\times mn$ stochastic
matrix. With this fact, the graph of $\bar{G}(t)$ is a directed
graph with $mn$ vertices. It is not difficult to see that
$\bar{G}(t)$ can be seen as an $n\times n$ \emph{block circulant
matrix} with blocks of size $m\times m$ where each row  block vector
is rotated one block to the right relative to the preceding row
block vector.

Let $\bar{\mathbb N}(t)$ be the graph of $\bar{G} (t)$. $\bar
{\mathbb{N}} (t)$ has the following properties.

\begin{lemma}\label{lemma:sign-arc}
For any $i,j\in \mathbf{n}$,  if $g_{ij}(t)=\alpha_q$ where $q\in
\underline{\mathbf{m}}$,  $\bar  {\mathbb{N}} (t)$ has an arc from
vertex $j+((p+q)\text{ mod }m)n$ to vertex $i+pn$ for $p\in
\underline{\mathbf{m}}$. In particular, $\bar {\mathbb{N}} (t)$ has
self-arcs at all $mn$ vertices.
\end{lemma}

\begin{lemma}\label{lemma:sign_path}
Suppose that $  \mathbb{N} (t)$ has a directed path from vertex $i$
to vertex $j$ with $i,j\in\mathbf{n}$. Then $\bar  {\mathbb{N}} (t)$
has a directed path from vertex $i$ to vertex $j+((m-q)\text{ mod
}m)n$ if the directed path from $i$ to $j$ in $  \mathbb{N} (t)$ has
a gain $\alpha_{q}$ for $q\in \underline{\mathbf{m}}$.
\end{lemma}

\begin{lemma}\label{lemma:path}
For a fixed $q \in \underline{\mathbf{m}}$, $\bar {\mathbb{N}} (t)$
has a directed path from vertex $i+pn$ to vertex $j+((m-q+p)\text{
mod } m)n$ with $i,j\in \mathbf{n}$ $p\in \underline{\mathbf{m}}$,
if and only if it has a directed path from vertex $i$ to vertex
$j+((m-q)\text{ mod }m)n$.
\end{lemma}

\begin{figure}[h]
\centerline{\includegraphics[height = 0.5in]{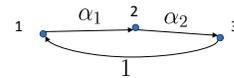}}
\centering \caption{Graph $\mathbb N$ associated with the sign set
$\{1, \alpha_1,\alpha_2\}$ } \label{gf:1}\end{figure}

\begin{figure}[h]
\centerline{\includegraphics[height = 1.9in]{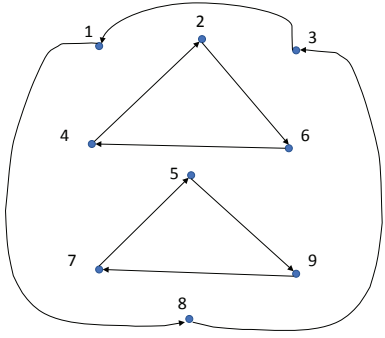}}
\centering \caption{Graph $\bar{\mathbb N}$}
\label{gf:2}\end{figure}

Look at the example below. For simplicity, self-arcs in the graphs
are eliminated. Fig.~\ref{gf:1} is the graph of $G$, which is a
three vertex graph associated with the gain set $\{1,
\alpha_1,\alpha_2\}$ where $\alpha_1=e^{\frac{2\pi}{3}j}$ and
$\alpha_2=e^{\frac{4\pi}{3}j}$. Correspondingly, the matrix $G$ for
system \eqref{eq:update_sys} associated with Fig.~\ref{gf:1} is
\[G=\begin{bmatrix} \frac{1}{2} & 0 & \frac{1}{2}\\ \frac{1}{2} \alpha_1 &\frac{1}{2} & 0
\\ 0 & \frac{1}{2} \alpha_2& \frac{1}{2}
\end{bmatrix}\] The matrix $\bar G$ for system \eqref{eq:expand_state} is
\[\bar{G}=\left[\begin{array}{lll|lll|lll} \frac{1}{2} & 0 &  \frac{1}{2} & 0& 0& 0&0&0&0\\ 0 & \frac{1}{2} &0 &\frac{1}{2}& 0& 0&0&0&0 \\   0& 0& \frac{1}{2}& 0&0&0&0&\frac{1}{2}& 0 \\ \hline 0&0&0& \frac{1}{2}& 0 & \frac{1}{2} & 0&0&0 \\ 0&0 &0 &0 &\frac{1}{2} & 0 & \frac{1}{2}& 0&0 \\ 0&\frac{1}{2}&0&0&0& \frac{1}{2}&0&0&0\\ \hline 0 &0&0&0&0&0 & \frac{1}{2}& 0& \frac{1}{2}\\ \frac{1}{2}&0&0 &0&0&0 &0 & \frac{1}{2}&0\\
0&0&0& 0&\frac{1}{2}& 0 & 0&0& \frac{1}{2}
\end{array}\right] \]
It is easy to see that $\bar{G}$ is a $3\times 3$ block circulate
matrix with blocks of size $3\times 3$. The graph $\bar{\mathbb{N}}$
of $\bar G$ is shown in Fig.~\ref{gf:2}.


\begin{proposition}\label{prop:balanced}
Suppose that the gain graph of $G(t)$ associated with the gain set
$\mathcal{G}$ is strongly connected and structurally $m$-balanced
with respect to a clustering vector $b\in \mathcal{I}$. Then, the
graph of $\bar{G}(t)$ consists of $m$ disjoint strongly connected
components of the same size, $n$.
\end{proposition}
\vspace{0.1in} \noindent{\bf Proof:} Since the graph of $G(t)$,
$\mathbb{N}(t)$, is structurally $m$-balanced with respect to a
clustering vector $b\in \mathcal I$, according to
Proposition~\ref{prop:set}, there exist $m$ disjoint vertex sets
$V_1,\;V_2,\;\ldots, \;V_m$ such that $\cup_{p=1}^mV_p=\mathbf{n}$
and for any vertex $i\in \mathbf{n}$, $i\in V_p$ if
$b(i)=\alpha_{p-1}$ for any $p\in \mathbf{m}$. For the $m$ disjoint
vertex sets $V_1,\;V_2,\;\ldots, \;V_m$,  $1\in V_1$, and all
semi-walks from vertex $i\in V_q$ to $j\in V_p$ have the same gain
$\alpha_{p-q}$ for any $p\geq q \in \mathbf{m}$. Let $V_{pq} $ for
any $p,q\in\mathbf{m}$ be a vertex set such that
\begin{equation}\label{eq:V}
V_{pq}=\{(p-1)n+i|\forall i\in V_q\}.
\end{equation}
 We get that $V_{pq}$ are disjoint for different $p,q$, and $\cup
_{p=1}^m\cup_{q=1}^mV_{pq}$ is the vertex set of $\bar{\mathbb N}$.
Note $V_{1q}=V_q$ for any $q\in \mathbf{m}$. Next we are going to
show that the following $m$ components for $p \in  \mathbf{m}$
 \begin{equation}\label{eq:C_p}
\mathcal C_p=\{V_{p,1}, V_{p-1,2},\ldots, V_{1,p}, V_{p+1,m},
V_{p+2,m-1},\ldots, V_{m,p+1}\}
\end{equation}
are disjoint. Moreover, each component is strongly connected and has
size $n$.

Since the size of $V_{pq}$ is same as the size of $V_q$. The size of
$\mathcal C_p$ is $n$. To begin with, we are going to show that any
two vertices in $\mathcal C_p$ are mutually reachable for $p\in
\mathbf{m}$. Since $\bar {G}$ is an $n\times n$ block circulate
matrix, if any two vertices in $\mathcal C_m$ are mutually
reachable, any two vertices in $\mathcal C_p$ for any $p\in
 \mathbf{m}$ are mutually reachable.  Now look at $\mathcal C_m=\{V_{m1}, V_{m-1,2},\ldots, V_{1m}\}$
Arbitrarily choose two nonempty elements of $\mathcal C_m$. Say
$V_{m-p+1,p}$ and $V_{m-q+1,q}$ where $1\leq p< q\leq m$. Due to the
definition, in graph $\mathbb N$, there is a path from a vertex $i
\in V_p$ to a vertex $j\in V_q$ with a gain $\alpha_{q-p}$.
According to Lemma~\ref{lemma:sign_path}, $\bar{\mathbb N}$ has a
directed path from vertex $i$ to vertex $j+(m-q+p)n$. Moreover,
according to Lemma~\ref{lemma:path}, $\bar{\mathbb N}$ has a
directed path from a vertex $i +(m-p)n$ which is in $V_{m-p+1, p}$
to a vertex $j+(m-q)n$ which is in $V_{m-q+1,q}$. Since the graph
$\mathbb{N}$ is strongly connected and structurally $m$-balanced, if
there  is a path from a vertex $i \in V_p$ to a vertex $j\in V_q$
with a gain $\alpha_{q-p}$, there must be a path from the vertex $j
\in V_q$ to a vertex $i\in V_p$ with a gain $\alpha_{m-q+p}$.
Similarly, we get the result that $\bar{\mathbb N}$ has a directed
path from a vertex $j +(m-q)n$ which is in $V_{m-q+1, q}$ to a
vertex $i+(m-p)n$ which is in $V_{m-p+1,p}$. That is, any two
vertices from $V_{m-p+1,p}$ and $V_{m-q+1,q}$ are mutually
reachable. Thus any  two vertices in $\mathcal C_p$ for $p\in
\mathbf{m}$ are mutually reachable.

Next, we prove that all the $m$ components are disconnected by
contradiction. Suppose there is a path from a vertex in $\mathcal
C_p$ to a vertex in $\mathcal C_q$ for $p< q$. Arbitrarily choose
two vertices $i$ and $j$ from $V_1$. Since we have shown that any
two vertices in $\mathcal C_p$ and $\mathcal C_q$ are mutually
reachable respectively, there is a path from a vertex $i+(p-1)n$ in
$ V_{p1}$ to a vertex $j+(q-1)n$ in $ V_{q1}$. According to
Lemma~\ref{lemma:path}, there is a path from a vertex $i$ to vertex
$j+(q-p)n$. This means that there is a path from vertex $i$ to
vertex $j$ with a gain $\alpha_{m-q+p}$. But both $i$ and $j$ belong
to $V_1$ which means that the path between these two vertices should
have gain 1. A contradiction. Thus all the $m$ components are
disconnected. \hfill $\qed$ \vspace{0.1in}

\begin{figure}[h]
\centerline{\includegraphics[height = 1.7in]{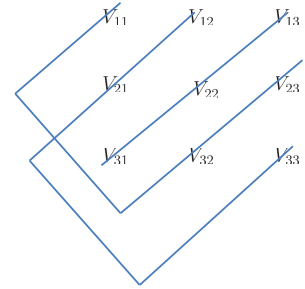}}
\centering \caption{Vertex sets of strongly connected components of
graph $\bar{\mathbb N}$ } \label{gf:3}\end{figure}

Look at the graph $\mathbb{N}$ in Fig.~\ref{gf:1}. $\mathbb{N}$ is a
strongly connected and structurally $3$-balanced graph. Here we can
get $V_1=\{1,3\}$, $V_2=\{2\}$, and $V_3=\emptyset$ where the
semi-walks from a vertex in  $V_1$ to a vertex in $V_2$ have gain
$\alpha_1$. Correspondingly, we get the expanded graph
$\bar{\mathbb{N}}$ as shown in Fig.~\ref{gf:2}.  For graph
$\bar{\mathbb N}$, we have $V_{11}=\{1,3\}$, $V_{12}=\{2\}$,
$V_{21}=\{4,6\}$, $V_{22}=\{5\}$, $V_{31}=\{7,9\}$, $V_{32}=\{8\}$,
and $V_{13}=V_{23}=V_{33}=\emptyset$. Three disjoint strongly
connected components of size $3$ are achieved as shown in
Fig~\ref{gf:3}. The first component consists of vertex $1$, $3$, and
$8$. The second component consists of vertex $2$ , $4$, and $6$. And
the last component consists of vertex $5$, $7$, and $9$.

\begin{proposition}\label{prop:unbalanced}
Suppose that the graph of $G(t)$ is strongly connected and
structurally unbalanced. Then, the graph of $\bar{G}(t)$  consists
of at most $\lfloor \frac{m}{2} \rfloor$ disjoint strongly connected
components, of at least size $2n$.
\end{proposition}
\vspace{0.1in} \noindent{\bf Proof:} Since the graph of $G(t)$,
i.e., $\mathbb{N}(t)$ is strongly connected and structurally
unbalanced, for a fixed vertex $i$ and any other vertex $j$ in
$\mathbb{N}(t)$, suppose there is a path from $i$ to $j$ with a gain
$\alpha_p$ and there is a path from $j$  to $i$ with  a gain
$\alpha_q$ such that $1\leq p+q<m$. According to
Lemma~\ref{lemma:sign_path}, there is a path from $i$ to $j+(m-p)n$
in graph $\bar{\mathbb N}$, and a path from $j$ to $i+(m-q)n$ in
$\bar{\mathbb N}$. Based on Lemma~\ref{lemma:path}, there is a path
from $j+(m-p)n$ to $i+((2m-p-q)\text{ mod }m)n$ in $\bar{\mathbb
N}$. Since $p+q< m$, $(2m-p-q)\text{ mod }m \neq i$ that is
$i+((2m-p-q)\text{ mod }m)n$ can not be vertex $i$. Based on
Lemma~\ref{lemma:path}, there is a path from $i+((2m-p-q)\text{ mod
}m)n$ to $j+((3m-2p-q)\text{ mod }m)n$. Repeat this procedure,
eventually there is a path from vertex $j+(((2k-1)m-kp-(k-1)q)
\text{ mod }m)n$ to $i+((2km-kp-kq)\text{ mod }m)n$ in $\bar{\mathbb
N}$ where $k$ is an integer which is great than 1. There exist $k$
such that $(2km-kp-kq)\text{ mod }m=0$. The easiest choice is to let
$k=m$. It means that there is a cycle starting from vertex $i$,
passing vertex $j+(m-p)n$, vertex $i+((2m-p-q)\text{ mod }m)n$
$\ldots$ and eventually ending with vertex $i$ again.  Since $j$ can
be any other vertex in $\mathbb N$, for a component in graph
$\bar{\mathbb N}$ consisting of agent $i$, it is strongly connected
and the size must be greater than $2n$. \hfill $\qed$ \vspace{0.1in}

Next we are going to prove the main results of this paper.

\textbf{Proof of Theorem~\ref{thm:1}:} (Sufficiency)
Since the sequence of neighbor graphs $
\mathbb{N}(1),\mathbb{N}(2),\ldots$ is repeatedly jointly strongly
connected, and repeatedly jointly structurally $m$-balanced with
respect to the clustering vector $b$, without loss of generality,
suppose there exists positive integers $p$ and $q$ such that each
finite sequence of graphs
$\mathbb{N}(q+kp),\mathbb{N}(q+kp+1),\ldots, \mathbb{N}(q+kp+p-1)$
is jointedly strongly connected and jointedly structurally
$m$-balanced with respect to the clustering vector $b$ for $k\geq
0$. Let \[\mathbb{H}(k)=\mathbb{N}(q+kp)\cup
\mathbb{N}(q+kp+1)\cup\ldots\cup \mathbb{N}(q+kp+p-1)\] Then
$\mathbb{H}(k)$ is strongly connected and structurally $m$-balanced
with respect to the clustering vector $b$. According to Proposition
~\ref{prop:set}, there exist $m$ disjoint vertex sets
$V_1,\;V_2,\;\ldots, \;V_m$ such that $\cup_{p=1}^mV_p=V$. Without
loss of generality, let $1\in V_1$, and all  walks from vertex $i\in
V_q$ to $j\in V_p$ have the same gain $\alpha_{p-q}$ for any $p\geq
q \in \mathbf{m}$. Now consider the expanded graph
$\bar{\mathbb{H}}(k)$ which is $\bar {\mathbb{H}}(k)=\bar
{\mathbb{N}}(q+kp)\cup \bar { \mathbb{N}}(q+kp+1)\cup\ldots\cup \bar
{\mathbb{N}}(q+kp+p-1)$. According to
Proposition~\ref{prop:balanced}, graph $\bar{\mathbb{H}}(k)$
consists of $m$ disjoint strongly connected components of the same
size $n$. $V_{pq}$ and $\mathcal{C}_p$ are defined the same as
Eq.~\eqref{eq:V} and \eqref{eq:C_p}. Each $\mathcal{C}_p$ is a
strongly connected component of size $n$. According to the result of
discrete-time linear consensus process \cite{reachingp1}, all the
vertices in $\mathcal{C}_p$ achieve consensus exponentially fast for
almost all initial conditions. From the structural of
$\mathcal{C}_p$ and system ~\eqref{eq:expand_state}, for any fixed
$p,q\in \mathbf{m}$, $i\in V_p ,\; j\in V_q$,
\[\lim_{t\rightarrow \infty} \alpha_{p-1}x_i(t)=
\lim_{t\rightarrow \infty} \alpha_{q-1}x_j(t) \neq 0.\] That is the
same as we say system~\eqref{eq:update_i} or ~\eqref{eq:update_sys}
achieves \textit{m-modulus consensus} corresponding to the
clustering vector $b$ exponentially fast for almost all initial
states .

(Necessity) Prove by contradiction. Suppose  the sequence of
neighbor graphs $ \mathbb{N}(1),\mathbb{N}(2),\ldots$ is repeatedly
jointly strongly connected, but not repeatedly jointly structurally
$m$-balanced with respect to a clustering vector $b$. Two scenarios
need to be considered.

First, if the sequence of neighbor graphs is  repeatedly jointly
structurally $m$-balanced with respect to another clustering vector
$b_1$, according to sufficiency we just proved
system~\eqref{eq:update_i} achieves m-modulus consensus
corresponding to $b_1$ not $b$.

Second, the sequence of the neighbor graphs is not repeatedly
jointly structurally $m$-balanced with respect to any clustering
vector. It means that either structurally unbalanced  graphs or more
than one class of structurally $m$-balanced graphs, or both appear
infinitely many times. Since the sequence of neighbor graphs $
\mathbb{N}(1),\;\mathbb{N}(2),\;\ldots$ is repeatedly jointly
strongly connected, there exist two positive integers $q$ and $p$
such that each finite sequence of graphs
$\mathbb{N}(q+kp),\mathbb{N}(q+kp+1),\ldots, \mathbb{N}(q+kp+p-1)$
is jointedly strongly connected for $k\geq 0$. Let
\[\mathbb{F}(k)=\mathbb{N}(q+kp)\cup \mathbb{N}(q+kp+1)\cup\ldots\cup
\mathbb{N}(q+kp+p-1)\] Then $\mathbb{F}(k)$ is strongly connected.
If a gain  graph $\mathbb{N}$ is structurally unbalanced, then any
finite sequence of gain graphs which contains $\mathbb{N}$ must be
jointedly structural unbalanced. If two gain graphs $\mathbb{N}_1$
and $\mathbb{N}_2$ are structurally $m$-balanced with respect to two
different clustering vectors $b_1$ and $b_2$ correspondingly,  then
any finite sequence of gain graphs which contains $\mathbb{N}_1$ and
$\mathbb{N}_2$ must be jointedly structural unbalanced. Since either
structurally unbalanced gain graphs or more than one class of
structurally $m$-balanced graphs, or both appear infinitely many
times, the graphs in the sequence $\mathbb{F}(k)$ will be
structurally unbalanced for infinitely many times. There must exist
two integers $n_1,n_2$ satisfying $1\leq n_1+n_2<m$ such that the
graphs in the sequence $\mathbb{F}(k)$, which has a path from $i$ to
$j$ with a gain $\alpha_{n_1}$ and
    a path from $j$ to $i$ with a gain $\alpha_{n_2}$,  appear infinitely many times.
 From the proof of Proposition~\ref{prop:unbalanced}, if $\mathbb{F}(k)$  has a path from $i$ to $j$ with a gain $\alpha_{n_1}$ and
    a path from $j$ to $i$ with a gain $\alpha_{n_2}$, the union of expanded graph
 $\bar {\mathbb{F}}(k)=\bar {\mathbb{N}}(q+kp)\cup \bar { \mathbb{N}}(q+kp+1)\cup\ldots\cup
\bar {\mathbb{N}}(q+kp+p-1)$ must have at most $\lfloor \frac{m}{2}
\rfloor$ disjoint strongly connected components, of at least size
$2n$. Moreover,
 vertex $i$,
vertex $j+(m-n_1)n$, vertex $i+((2m-n_1-n_2)\text{ mod }m)n$ must
belong to one strongly component. According to \cite{cut},  state
$z(i)$ and $z(i+((2m-n_1-n_2)\text{ mod }m)n)$ would achieve
consensus asymptotically which means that $z(i)$ would converge to
zero asymptotically fast. Since $i$ is randomly chosen, $z$
converges to zero asymptotically fast. Thus the sequence of the
neighbor graphs is repeatedly jointedly structurally $m$-balanced
with respect to the clustering vector $b$. \hfill $\qed$

\vspace{0.1in}

\textbf{Proof of Theorem~\ref{thm:2}:}
 Since the sequence of neighbor graphs   $ \mathbb{N}(1),\mathbb{N}(2),\ldots$ with the same gain set $\mathcal{G}$  is repeatedly jointly strongly connected and repeatedly jointly structurally unbalanced,  system~\eqref{eq:update_sys}
converges to zero asymptotically fast for almost all initial
conditions based the analysis of necessity of theorem~\ref{thm:1}.
\hfill $\qed$

\section{Conclusion}

In this paper, a generalized discrete-time Altafini model  over
time-varying gain graphs, in which  the arcs are assigned complex
numbers from a cyclic group whose order determines the maximum
possible number of clusters, has been studied through a graphical
approach. Necessary and sufficient conditions for exponential
convergence of the system with respect to nonzero limit states have
been established under the assumption of repeatedly jointly strong
connectivity. A sufficient condition for asymptotic consensus at
zero has also been provided. The results in this paper can be
extended to the case where the gains of the neighbor graph are the
elements of a finite abelian group. Necessary and sufficient
conditions for exponential convergence at zero of the system will be
studied in the future. The time-varying case without the strong
connectivity assumption is another direction for future research.
\bibliographystyle{unsrt}
\bibliography{clustering,social}
\end{document}